# Probing the reciprocal lattice associated with a triangular slit to determine the orbital angular momentum for a photon

W. C. SOARES,[1,2,*] A. L. MOURA,[1] ASKERY CANABARRO,[1,3] E. DE LIMA,[1] J. H. LOPES,[1] E. J. S. FONSECA,[4] M. L. FELISBERTO,[5] B. DE LIMA BERNARDO,[6] J. M. HICKMANN,[7] AND S. CHÁVEZ–CERDA[8]

[1]*Grupo de Física da Matéria Condensada, Núcleo de Ciências Exatas – NCEX, Campus Arapiraca, Universidade Federal de Alagoas, Arapiraca, Alagoas 57309-005, Brazil*
[2]*Grupo de Informação Quântica do Sul – GIQSul, Departamento de Física, Universidade Federal de Catarina, Florianópolis, Santa Catarina 88040-900, Brazil*
[3] *International Institute of Physics, Federal University of Rio Grande do Norte, 59070-405, Brazil*
[4]*Instituto de Física, Universidade Federal de Alagoas, Maceió, Alagoas 57061-970, Brazil*
[5]*Universidade Federal de Alagoas, Campus do Sertão, Delmiro Gouveia, Alagoas 57480-000, Brazil*
[6] *Universidade Federal da Paraíba, João Pessoa, Paraíba, 58051-900, Brazil*
[7] *Universidade Federal do Rio Grande do Sul, Porto Alegre, Rio Grande do Sul, 91501-970, Brazil*
[8]*Instituto Nacional de Astrofísica, Óptica y Eletrónica, Luiz Enrique Erro No. 1, Tonantzintla, Puebla 72840, Mexico*
*\*willamys@fis.ufal.br*

**Abstract:** The orbital angular momentum conservation of light reveals different diffraction patterns univocally dependent on the topological charge of the incident light beam when passing through a triangular aperture. It is demonstrated that these patterns, which are accessed by observing the far field measurement of the diffracted light, can also be obtained using few photon sources. In order to explain the observed patterns, we introduce an analogy of this optical phenomenon with the study of diffraction for the characterization of the crystal structure of solids. We demonstrate that the finite pattern can be associated to the reciprocal lattice obtained from the direct lattice generated by the primitive vectors composing any two of the sides of the equilateral triangular slit responsible for the diffraction. Using the relation that exists between the direct and reciprocal lattices, we provide a conclusive explanation of why the diffraction pattern of the main maxima is finite. This can shed a new light on the investigation of crystallographic systems.

## 1. Introduction

Photons are the main information carriers in quantum optics. In fact, single and entangled photons have played an important role in the development of quantum computation and quantum information technologies [1]. Quantum bits (qubits) of information can be encoded in different degrees of freedom as, for instance, the single photon energy, polarization, linear momentum and orbital angular momentum (OAM) state. Recently, the OAM degree of freedom of the photon has received a lot of attention for providing a discrete high-dimensional quantum space [2,3]. It has been demonstrated that this new degree of freedom allows the preparation of single and entangled photons in a superposition of *n* orthogonal quantum states (i.e. qudits instead of qubits) [4-7]. As a consequence, a single photon can carry a big amount of information encoded on its OAM state. In this regard, the study of photon's OAM has provided novel counterintuitive examples on the relationship between the quantum and the classical regime [8,9]. In this form, it is essential to develop ways of measuring the OAM states of light from both the quantum and classical regimes.

There are already several methods to determine the OAM of light with many photons in the same mode. Usually, the beam's OAM is experimentally determined by interfering the beam possessing OAM with a reference plane wave [10], or with its mirror image [11]. More recently, new techniques for obtaining the OAM state have been reported. Some of them are related with a direct measurement of the wavefront of the beam [12], and the interference with a double-slit [13]. The relationship between OAM states and diffraction through triangular apertures has been studied and has become very popular among experimentalists [14-17].

An important issue in quantum communication processing relies in the ability of measuring a quantum state. An efficient, precise and easy technique to determine an OAM eigenstates for any value of *m* is still challenging. In some experiments, an arrangement of holograms, single mode fibers and single photon counting module detector has been used to determine the OAM state [5-8]. A Laguerre-Gaussian ($LG_{m,p}$) mode can be transformed into a Gaussian mode by using holographic techniques. Notice that for each measurement of a specific OAM quantum state it is necessary an appropriate hologram, limiting this method to measure only one particular state. A more complex hologram can be projected for sorting OAM modes [18, 19]. In this context, the development of accurate techniques to measure high orders states of the OAM of a single photon is an important issue that must be strongly investigated due to the importance of their applications for quantum information processing.

In this paper, we present a connection between the diffraction phenomena of photons endowed with OAM and the formation of an optical lattice in the far field plane. The same diffraction pattern is also obtained in the regime of a classical optical field carrying OAM. The physical description of this phenomenon is formulated with basis on the study of crystal structures in the theory of solid state physics. Specifically, this far field diffraction pattern is associated to the reciprocal lattice of a direct lattice, whose primitive vectors represent the sides of the triangular slit used to produce the diffraction pattern. Due to the OAM conservation, only finite reciprocal lattices are exhibited allowing the determination of the OAM state of the photons.

## 2. Finite Diffraction Lattice

To start this section, we describe the quantum state of photons in terms of the OAM spatial modes. In the paraxial approximation, the $LG_{m,p}$ modes constitute a complete infinite-dimensional basis. Alternatively, it has been demonstrated that the $LG_{m,p}$ modes can be identified as the eigenstates of a quantum OAM operator possessing eigenvalues $m\hbar$ [20,21], i.e., $L_z |m,p\rangle = m\hbar |m,p\rangle$, where $|m,p\rangle$ represents the photon state prepared in the $LG_{m,p}$ mode. Such $LG_{m,p}$ modes are characterized by two integer numbers $p$ and $m$, where $p$ represents the radial mode and $m$ determines the dependence of the modes with the azimuthal phase in the form $\exp(im\varphi)$, being also referred to as the topological charge. This term is what defines an optical vortex that is the one associated to the OAM of the photon. Let us now discuss the OAM of a beam and the resulting diffraction pattern by a triangular thin slit aperture using some elements from the classical perspective for this problem and its relation to diffraction by crystals.

It is well known that the total angular momentum density of an optical beam is given by

$$\mathbf{j} = \mathbf{r}_\perp \times \mathbf{p}, \tag{1}$$

where $\mathbf{p}$ is the linear momentum density vector given by $\mathbf{p} = \varepsilon_0 \mathbf{E} \times \mathbf{B}$, and $\mathbf{r}_\perp$ is the transverse coordinate at the aperture plane. $\mathbf{E}$ and $\mathbf{B}$ are respectively the electric and magnetic fields, and $\varepsilon_0$ the vacuum permittivity. Considering the situation sketched in Fig. 1(a), it is easy to see that the OAM density z component can be written as $j_z = r_\perp p_\phi$, where $p_\phi$ is the azimuthal component of the linear momentum **p**. For such an incident beam, the OAM per photon is given by $j_z = m\hbar$ and the local transverse linear momentum is $p_\phi = \hbar k_\perp$, where $k_\perp$ is the transverse wavevector of the light beam, that defines the transverse wavelength $\lambda_\perp$ at a given plane perpendicular to the z-axis, and $\hbar$ is the reduced Planck's constant. By elimination of $\hbar$ from these two equations we get $r_\perp k_\perp = m$. For convenience we define the normalized variable $r_\perp = r_\perp'/2\pi$ to get

$$k_\perp r_\perp' = 2\pi m. \qquad (2)$$

This relation is satisfied for any light beam with OAM and its relevance will be clear below.

In order to understand the diffraction pattern of single photons by a triangular slit we resource to the theory of diffraction by crystals. For this purpose, we will consider the equilateral triangular slit as a unit cell of a two-dimensional Bravais triangular lattice defined by the vectors $\mathbf{a}_1$ and $\mathbf{a}_2$ such that $|\mathbf{a}_1| = |\mathbf{a}_2|$. The corresponding reciprocal lattice is created with the vectors $\mathbf{b}_1$ and $\mathbf{b}_2$ that satisfy the condition $\mathbf{a}_i \cdot \mathbf{b}_j = 2\pi\delta_{ij}$ where $\delta_{ij}$ is the Kronecker's delta [22]. This condition implies that the direct lattice and the reciprocal lattice are formed by mutually orthogonal vectors as shown in Fig. 1(b).

Any point in the direct lattice space is represented by $\mathbf{r}_\perp' = u_1\mathbf{a}_1 + u_2\mathbf{a}_2$ and the points in the reciprocal space is characterized by the set of vectors $\mathbf{k}_\perp = v_1\mathbf{b}_1 + v_2\mathbf{b}_2$, where the coefficients $u_i$ and $v_j$ are integer numbers. Since $\mathbf{r}_\perp'$ is in the Bravais lattice and $\mathbf{k}_\perp$ belongs to the reciprocal lattice we have the Laue condition [22]

$$\exp(i\mathbf{k}_\perp \cdot \mathbf{r}_\perp') = 1 \qquad (3)$$

implying that the dot product in the argument is an integer multiple of $2\pi$. this is $\mathbf{k}_\perp \cdot \mathbf{r}_\perp' = 2\pi\eta$, where $\eta$ can be any integer number. Explicitely, Eq. (3) becomes

$$u_1 v_1 + u_2 v_2 = \eta. \qquad (4)$$

The direct lattice under investigation is only comprised by one single cell with the vectors $\mathbf{a}_1$ and $\mathbf{a}_2$, therefore, either of the coefficients $u_i$ can be equal to zero or one to define de vertices of the Bravais triangular unit cell. Recalling that the modulus of the product of the vectors $\mathbf{r}_\perp'$ and $\mathbf{k}_\perp$ are related to the beam topological charge $m$ by Eq. (2) and assuming, without loss of generality, that $u_1 = u_2 = 1$ in Eq. (4) we obtain

$$v_1 + v_2 = m, \qquad (5)$$

that sets the limit of the maximum value that cant take the integer $\eta$ such that Eq. (3) be satisfied. Thus, each the $v_j$ is restricted to have values from 0 to $m$ in such a way that their sum does not exceed the maximum value of $m$ (notice that if $m$ is negative also must be the $v_j$'s up to the minumum value of $-|m|$). The resulting diffraction pattern is then governed by the Laue condition that states that constructive interference will occur whenever the change in wave vector $\mathbf{k}_\perp = v_1\mathbf{b}_1 + v_2\mathbf{b}_2$ is a vector of the reciprocal lattice. In this way, the diffraction pattern must reveal the corresponding part of the reciprocal lattice associated with the topological charge $m$.

Since the sum in Eq. (5) is restricted to the value of $m$ then the reciprocal Bravais lattice is finite. By increasing the value of $m$, new portions of the reciprocal lattice should be unveiled. In Fig. 1(c), we can see an example where the construction of the reciprocal Bravais lattice is shown for an equilateral triangular unit cell for $m = 3$. Notice that the reciprocal vectors also define the orientation of the triangular diffraction pattern. One final remark, the analisys has been done for a vortex wave front disregarding the amplitude of the light beam. The theory developed above will be satisfied for vortex beams whose intensity is approximated by the simplest representation of a vortex, namely $(x\pm iy)^m = r^m e^{im\phi}$. For light beams like Laguerre-Gauss beams or Bessel beams this occurs when the triangular slit aperture is centered inside the first ring of their intensity pattern, atmost circunscribed.

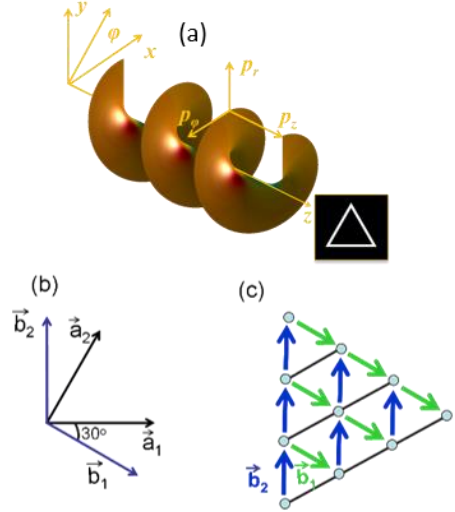

Fig. 1: (a) Schemactical representation of diffraction of a vortex light beam by an equilateral triangular slit. (b) Definition of the direct and reciprocal lattice vectors corresponding to a triangular lattice. (c) Triangular lattice formed with photons with $m=3$ of topological charge.

## 3. Results

Let us consider the diffraction problem where photons endowed with OAM are scattered by an equilateral triangular slit forming a diffraction pattern at the Fourier plane. The photodetection rate $N(\rho)$, transmitted by the triangular slit to a very small detector located at the Fourier plane is proportional to the second order correlation function of the field [23]

$$N(\rho) \propto \left\langle \hat{E}_T^-(\rho)\hat{E}_T^+(\rho) \right\rangle, \quad (6)$$

where $\hat{E}_T^-(\rho)$ and $\hat{E}_T^+(\rho)$ are the negative and positive frequency components, respectively, of the electric field operator measured at the Fourier plane, and $\rho$ is the transverse position vector at the detector plane.

The electrical field operator is obtained by making an analogy with the classical calculation of the electric field transmitted through an object when the angular spectrum is known [24]. The transmitted electrical field operator at the Fourier plane is written as

$$\hat{E}_T^+(\rho) \propto \int d\mathbf{q}\,\hat{a}(\mathbf{q})\Delta(\mathbf{q})e^{i\mathbf{q}\cdot\rho} \quad . \quad (7)$$

In this equation, the operator $\hat{a}(\mathbf{q})$ annihilates a photon with a transverse wavevector $\mathbf{q}$ and $\Delta(\mathbf{q})$ is the Fourier transform of the transmission function that represents the object; in our case the equilateral triangular slit. The operators $\hat{E}_T^-(\rho)$ and $\hat{E}_T^+(\rho)$ contain all information about the presence of the optical elements in the path of the propagation from the slit to the detector.

Following along the lines of Ref. [25], and restricting to the OAM contribution only, i.e., neglecting the spin contribution, the state of a photon in a light beam propagating according to the paraxial wave equation can be expressed as

$$|\psi\rangle = \int d\rho\, \vartheta_{m,p}(\rho)|\rho\rangle .\qquad(8)$$

This equation represents a single photon state in the paraxial mode with indices $m$ and $p$, and $\vartheta_{m,p}(\rho)$ corresponds to the transverse spatial wave function.

It is convenient to express the paraxial function, $\vartheta_{m,p}(\rho)$, in terms of its angular spectrum, $v_{m,p}(\mathbf{q})$, via the Fourier transform

$$\vartheta_{m,p}(\rho) \propto \int d\mathbf{q}\, v_{m,p}(\mathbf{q}) e^{i\mathbf{q}\cdot\rho} .\qquad(9)$$

By manipulating the equations above, we obtain

$$N(\rho) \propto \left| \int d\mathbf{q}\, v_{m,p}(\mathbf{q}) \Delta(\mathbf{q}) e^{i\mathbf{q}\cdot\rho} \right|^2 .\qquad(10)$$

This equation allows us to numerically calculate the probability distribution at the Fourier plane of a single photon prepared in an OAM state. This result corresponds to the far field diffraction pattern of the photons in the $LG_{m,p}$ mode scattered by an equilateral triangular slit. This probability distribution has a triangular pattern, as showed in Fig. 2. It is important to call attention to the fact that this pattern, which is formed after many detections of the photons, also corresponds the diffraction pattern formed by a classical optical field prepared and submitted to similar conditions. The present results were obtained by evaluating Eq. (10) for $m$ ranging from $1$ to $6$, with $p = 0$. Interesting enough, from these patterns we can observe a direct relationship between the number of maxima that shows up at each diffraction pattern with its corresponding value of $m$. Note that each side of the triangular diffraction patterns has $m+1$ bright spots. We can observe that the value of $m$ is directly related to the number of interference maxima along the external side of each triangular diffraction pattern. Indeed, the topological charge of the measured mode is $m = N - 1$, where $N$ is the number of maxima along any of the sides of the triangular pattern.

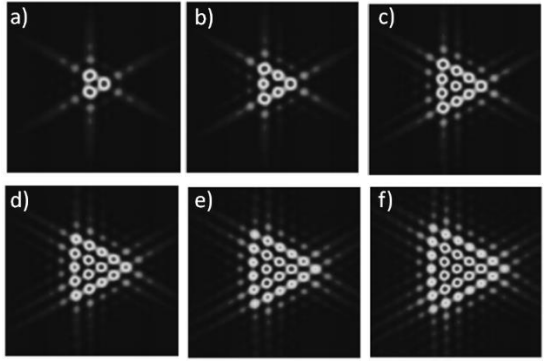

Fig. 2: Probability distribution at the Fourier plane formed after the detection of many photons. Numerical solution of Eq. (10) using Laguerre-Gauss modes diffracted by the equilateral triangular slit. The values of $m$ are (a) $m=1$, (b) $m=2$, (c) $m=3$, (d) $m=4$, (e) $m=5$, (f) $m=6$.

From a crystallography viewpoint, a similar diffraction problem appears when an X-ray beam illuminates, for example, a two-dimensional crystal structure. An integral similar to that shown in Eq. (10) is used to determine the scattered wave amplitude of X-rays [26], establishing a relation between the vectors in the direct and reciprocal spaces. In this case, $\Delta(\mathbf{q})$ has the meaning of the Fourier transform of the electron number density function and $\mathbf{q}$ corresponds to the reciprocal lattice vectors. The similarity between these integrals leads us to make an analogy between the present case of the diffraction of photons endowed with OAM by an aperture and the ideas from crystallography theory.

Calvo et al. [27], following the formalism of Ref. [25], pointed out that even though it has been carried out the paraxial quantization description using the $LG_{m,p}$ modes, other paraxial modes or even nonparaxial modes could be used as well. In this regards, we notice that Eq. (10) also supports high orders of Bessel modes, other families of modes that can be used to describe a light beam possessing OAM. Bessel modes are exact solution of the Helmohltz equation [28]. Fig. 3 illustrates the probability distribution for $m=3$ and $m=-3$ using Bessel modes. In this case, due to the well-known multi-ring structure of Bessel modes, the configuration we used was such that the smallest ring completely illuminates the triangular slit. We observe that by changing the sign of $m$, the orientation of the diffraction pattern also changes, therefore, allowing the determination the sign of the photon OAM in a practical way. In comparing Fig. 2(c), using the $LG_{m,p}$ modes, with Fig. 3(a), using Bessel modes, no significative difference is observed. This fact implies that the diffraction pattern depends only on the photon OAM and it is basically independent of the radial amplitude distribution. Therefore, the method presented in this paper to measure the photon OAM is quite general and can be used for any OAM mode.

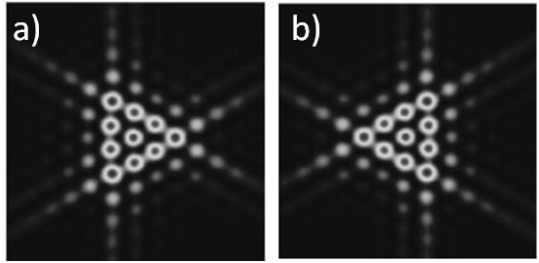

Fig. 3: Effect of the change of the sign of $m$ using Bessel modes in Eq. (10). The topological charges are (a) $m=3$ and (b) $m=-3$.

The experimental setup is sketched in Fig. 4. An Argon laser operating at *514 nm* illuminates a computer-generated hologram to produce high order $LG_{m,p}$ beams [29]. Using a pinhole, we select a well-defined $LG_{m,p}$ mode with $m=3$ and $p=0$. To reach the single photon level we used the same approach as that described in Ref. [32]; the power of this particular mode was strongly attenuated to produce an optical flux with a maximum of approximately 1500 photons/s. We have also selected a $LG_{m,p}$ mode with $m=-3$ and $p=0$. An equilateral triangular slit with side and slit length of $3.51\,mm$ and $0.4\,mm$, respectively, was placed in the photon's path. A lens with $300\,mm$ of focal length was placed after the triangular slit to generate the far field diffraction pattern at the focal plane, where the scattered photons were collected by a single mode fiber (SMF). The scattered patterns were obtained by scanning the fiber tip and recording the single photon counts (Perkin Elmer SPCM – AQR). In some measurements, the mirror just before the triangular slit was replaced by a pentaprism to change the sign of $m$ [29].

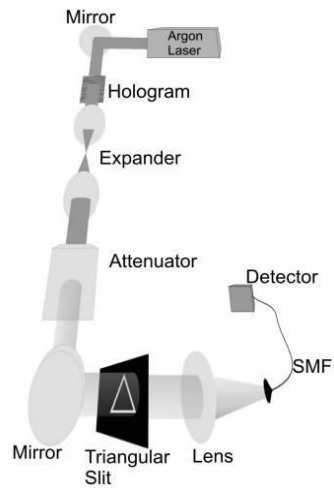

Fig. 4: Sketch of the experimental setup. The light from an Argon Laser acquires an specific OAM state after passing through a computer-generated hologram, before having its intensity strongly attenuated. The photons of this optical field illuminates the triangular slit before being detected, after which a lens and a single mode fiber (SMF) are used for the measurement to materialize in the far-field diffraction limit.

Fig. 5 shows the experimental (a, b) and the theoretical (c, d) results of the far field diffraction patterns for a single mode of the LG beam after many photons have been recorded. The far field diffraction patterns were measured by scanning a mono mode fiber tip in a matrix 6x6 mm. The maximum number of single counts per second was approximately 1500. By counting the number $N$, we can determine in which OAM state the photon was supposed to be. In our case, we have $N = 4$ which implies $m = 3$, in agreement with the OAM state of the photons diffracted. For $m = -3$ the orientation of the interference pattern changes as shown in Fig. 2 (b). Furthermore, this approach also allows us to understand the probability distribution at the Fourier plane of a single photon prepared in a OAM state for $m < 0$. We can understand this result by noting that $m < 0$ implies $v_1 < 0$ and $v_2 < 0$ in Eq. (5), therefore, the orientation of the reciprocal lattice should change as well. As we can observe in Fig. 5, the experimental results are in good agreement with the theory.

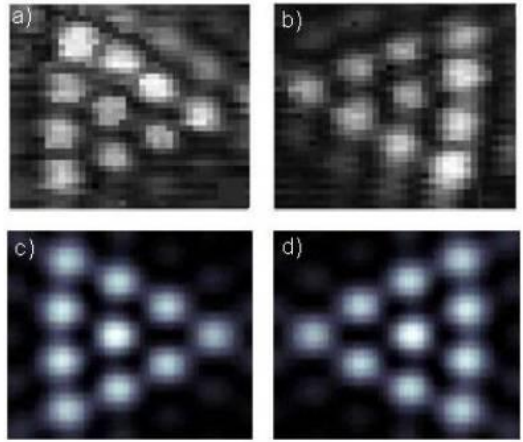

Fig. 5: Experimental results of the measurements of the light OAM in the after the measurement of several photons for (a) $m = 3$ and (b) $m = -3$. (c) and (d) show the edited images to evidence the contrast of the patterns shown in (a) and (b), respectively.

The method presented here works for determining any OAM mode corresponding to different values of $m$. For each value of $m$ we have a different probability distribution, allowing thus to precisely determine the photon's OAM. In fact, this method is very simple to be implemented and the results are obtained in a direct way without the necessity to change the experimental setup for different values of $m$ [14,17]. It is worthwhile to mention that probability distributions similar to the ones shown here can also be generated by an equilateral triangular aperture [15]. However, the diffraction pattern is better defined using the equilateral triangular slit because only the photons with the wavevectors defined by the slit width contribute to the reciprocal lattice formation, resulting in a better resolution between the consecutive maxima. Even though there is a relationship between the value of $m$ and the profile of the diffraction pattern [27-34], due to the symmetry properties, the equilateral triangle is the only configuration that produces diffraction patterns that allow to determine in a unambiguous way both the value and the signal of $m$. It is important to point out the expected behavior of the present experiment studied here for the case of photons with fractional and mixed OAM states. In the first case, since the helical wavefront characterizing the longitudinal phase profile is unstable [35], we do not expect a stationary diffraction pattern. In the second case, each photon has a set of classical probabilities of being measured in different OAM states, which renders an overlap of the corresponding diffraction patterns each with different intensities.

## 4. Conclusions

In conclusion, we theoretically and experimentally demonstrated the generation of finite optical lattices by means of the diffraction of light carrying OAM through an equilateral triangular slit. We established an analogy of this problem with that of diffraction in solid state physics, where important applications may arise from this idea. The obtained optical lattices correspond to specific regions, constrained by the topological charge of the incident photons, of the reciprocal lattice associated with a Bravais lattice, whose unit cell shape is triangular. The present ideas may also apply to other slit shapes like the square slit. The present findings addressing optical lattices may give rise to an alternative method of investigating the structure of solid crystals, once we they provide an enlarged version of these atomic complex systems, enabling a visual perspective. Although we have used the Laguerre-Gaussian modes, the present results can also be applied to other families of beams endowed with OAM like Bessel beams. A generalization of our results to periodic structures like atomic, photonic and plasmonic crystals and even to quasi-crystal structures could reveal unsuspected facets of the diffraction of photons carrying high-order OAM. It also add a new puzzle to the quantum Young's double slit experiment using vortical single photon sources or particles like electrons, neutral atoms, even complex fullerene molecules considering three slits arranged in a triangular shape.

**Funding.** We acknowledge with thanks the support from Coordenação de Aperfeiçoamento de Pessoal de Nível Superior (CAPES); Conselho Nacional de Desenvolvimento Científico e Tecnológico (CNPq); Fundação de Amparo à Pesquisa do Estado de Alagoas (FAPEAL); Instituto Nacional de Ciência e Tecnologia de Informação Quântica (INCT-IQ). WCS thanks P.H. Souto Ribeiro for discussion and comments on the theoretical results. AC acknowledge the John Templeton Foundation via the Grant Q-CAUSAL No. 61084, the Serrapilheira Institute (Grant No. Serra-1708-15763), the CNPq via Grant No. 423713/2016-7, also UFAL for a paid license for scientific cooperation at UFRN.

**Disclosures.** The authors declare no conflicts of interest.